\documentclass[aps,prc,twocolumn,nofootinbib,nobibnotes,floatfix,amsfonts,pdftex,superscriptaddress,longbibliography]{revtex4-1}

\usepackage{graphicx}  
\usepackage{amsmath, amssymb, bm, cancel}   

\usepackage{color}

\usepackage[unicode=true,
  linktocpage,
  linkbordercolor={0.5 0.5 1},
  citebordercolor={0.5 1 0.5},
  colorlinks=true,
  linkcolor=blue,
  citecolor=blue,
  urlcolor=blue]{hyperref} 

\newcommand{\threejm}[6]{ \begin{pmatrix}
  #1 & #2 & #3 \\
  #4 & #5 & #6 
 \end{pmatrix}}

 \newcommand{\nopieft}{\cancel{\pi}\mathrm{EFT}}

\begin{document}

\title{Neutrino-deuteron scattering: Uncertainty quantification and new $L_{1,A}$ constraints} 

\date{\today}

\author{%
Bijaya Acharya}
\email{acharya@uni-mainz.de}
\affiliation{Institut f\"{u}r Kernphysik and PRISMA$^+$ Cluster of Excellence,
  Johannes Gutenberg-Universit\"{a}t Mainz, 55128 Mainz, Germany}

\author{%
Sonia Bacca}
\email{s.bacca@uni-mainz.de}
\affiliation{Institut f\"{u}r Kernphysik and PRISMA$^+$ Cluster of Excellence,
  Johannes Gutenberg-Universit\"{a}t Mainz, 55128 Mainz, Germany}
  
\begin{abstract}
We study neutral- and charged-current (anti)neutrino-induced dissociation of the deuteron at energies from threshold up to 150 MeV by employing potentials, as well as one- and two-body currents, derived in chiral effective field theory ($\chi$EFT).
We provide uncertainty estimates from $\chi$EFT truncations of the electroweak current, dependences on the $\chi$EFT cutoff, and variations in the pool of fit data used to fix the low-energy constants of $\chi$EFT. At 100~MeV of incident (anti)neutrino energy, these uncertainties amount to about 2-3\% and are smaller than the sensitivity of the cross sections to the single-nucleon axial form factor, which amounts to $5\%$ if one varies the range of the nucleon axial radius within the bands determined by recent lattice quantum chromodynamics evaluations and phenomenological extractions. We conclude that a precise determination of the nucleon axial form factor is required for a high-precision calculation of the neutrino-deuteron cross sections at energies higher than 100~MeV.
By matching our low-energy $\chi$EFT results to those of pionless effective field theory ($\nopieft$), 
we provide new constraints for the counterterm $L_{1,A}$ that parametrizes the strength of the axial two-body current in $\nopieft$. We  obtain a value of $4.9^{+1.9}_{-1.5}~\mathrm{fm}^3$ at renormalization scale set to pion mass, which is compatible with, albeit narrower than, previous experimental determinations, and comparable to a recent lattice quantum chromodynamics calculation.
  
\end{abstract}

\maketitle

\section{Introduction}

Many fundamental questions in particle physics, astrophysics and cosmology are inextricably linked with neutrino properties and their interactions with nuclei. With the entry of neutrino experiments into an era of precision measurements, a reliable theoretical treatment of the scattering of neutrinos with nuclei that constitute the detector material is one of the most important challenges for nuclear physicists~\cite{Benhar:2015wva}. Precise theoretical calculations were first performed for light nuclei. Predictions for the (anti)neutrino-deuteron ($\bar\nu/\nu$-$d$) scattering cross sections by Nakamura {\it et al.}~\cite{Nakamura:2000vp,Nakamura:2002jg} marked a significant improvement over prior works (reviewed in Ref.~\cite{Kubodera:1993rk}) and played a crucial role in the analysis of experiments that led to the confirmation of neutrino oscillations~\cite{Ahmad:2002jz,Ahmed:2003kj}. These phenomenological calculations were based on the conventional meson-exchange model of nuclear interactions and weak currents. Shen {\it et al.}~\cite{shen} refined the modeling of the currents and extended the approach of Nakamura {\it et al.}~to neutrino energies up to the GeV scale. Efforts to extend these calculations to heavier nuclei are also under way. Breakup reactions of $^3$H and $^{3,4}$He were calculated in coordinate space using the method of hyperspherical harmonics in Refs.~\cite{Gazit:2004sp,Gazit:2007jt,OConnor:2007kup}, and $^{2,3}$H and $^{3}$He were treated in the momentum-space Faddeev formalism in Refs.~\cite{Golak:2018qya,Golak:2019fet}. The neutral weak responses of $^{4}$He~\cite{Lovato:2015qka} and $^{12}$C \cite{Lovato:2014eva,Lovato:2015qka,Lovato:2017cux} were studied using the Green's function Monte Carlo method. Inclusive $\nu$-$^{12}$C and $\nu$-$^{16}$O cross sections have been calculated using the correlated basis functions and self-consistent Green's function methods in Ref.~\cite{Rocco:2018mwt}. These studies have generally been motivated by the composition of the detector in past, present and future neutrino experiments such as SNO (heavy water), MiniBooNE (mineral oil), T2K/T2HK (water), and DUNE (liquid $^{40}$Ar).

The $\bar\nu/\nu$-$d$ cross sections have also been calculated in effective field theories, which provide a description of the scattering at low energies that correspond to a typical momentum scale $Q$ which is smaller than a breakdown momentum scale, $\Lambda_b$. The nuclear Hamiltonian and couplings to external electroweak sources are systematically constructed as 
perturbative expansions in $Q/\Lambda_b$ with controlled uncertainties. The expansion coefficients are functions of undetermined parameters called low-energy constants (LECs) which are usually fixed by fitting to experimental data. Pionless effective field theory ($\nopieft$), which has the nucleons as the only hadronic degrees of freedom, was applied to $\bar\nu/\nu$-$d$ scattering in Ref.~\cite{Butler:2000zp}. The results of prior phenomenological calculations were well reproduced for neutrino energies within the domain of convergence of the $\nopieft$ expansion, modulo fitting of a single undetermined LEC which is conventionally referred to as $L_{1,A}$. Recently, Baroni and Schiavilla~\cite{Baroni:2017gtk} performed the first calculation of $\bar\nu/\nu$-$d$ scattering in chiral effective field theory ($\chi$EFT), which uses nucleons and pions as effective degrees of freedom. Employing currents and interactions up to high orders in the $\chi$EFT expansion, Ref.~\cite{Baroni:2017gtk} obtained results that were consistent with, albeit 1-2\% larger than, the phenomenological calculations of Refs.~\cite{Nakamura:2000vp,shen}. 

In this work, we study the inelastic $\bar\nu/\nu$-$d$ scattering process in $\chi$EFT with several goals that are different from previous works. We set up an independent framework to express the $\chi$EFT operators as multipole expansions and then compare the various sources of uncertainties in the $\bar\nu/\nu$-$d$ cross section calculation. In addition to the approach in Ref.~\cite{Baroni:2017gtk} of fixing the potentials at a high $\chi$EFT order and examining the order-by-order contributions of the electroweak current operator, we also employ the NNLO$_\mathrm{sim}$ family of interactions~\cite{Carlsson:2015vda,Andreas} comprising 42 different $\chi$EFT potentials. 
These  potentials are all derived up to the third order in the $\chi$EFT expansion but span seven different values of regulator cutoffs and six different truncations of the maximum scattering energy in the world database of nucleon-nucleon (NN) scattering cross sections that were used to partly constrain the LECs. This allows for a more complete treatment of uncertainties. Furthermore, we investigate the sensitivity of the cross sections to variations in the nucleon axial radius within the uncertainties of recent lattice quantum chromodynamics (lattice QCD) evaluations and model-independent extractions, which are much larger than conventional error estimates obtained by assuming a dipole form factor, and compare this to $\chi$EFT uncertainties. Finally, by using our $\chi$EFT results as input, we constrain the value of the LEC $L_{1,A}$, which is a major source of uncertainty in $\nopieft$ calculations of nuclear weak processes such as the proton-proton fusion reaction occurring in our sun.

This paper is organized as follows. Section~\ref{sec:theory} briefly reviews the theory that relates the $\bar\nu/\nu$-$d$ cross section to electroweak response functions calculated from $\chi$EFT interactions and currents in a multipole-decomposition framework. In Sec.~\ref{sec:results}, we present the numerical results for the dissociation cross sections and discuss their implications. A brief summary and outlook are presented in Sec.~\ref{sec:conclusion}.

\section{Theory}
\label{sec:theory}

The cross section for $\bar\nu/\nu$-$d$ scattering off the deuteron follows from Fermi's golden rule (in natural units) as 
\begin{equation}
\label{eq:cross_sec_schematic}
 \sigma = \sum_h\int_\Phi \, \vert \langle f \vert \hat H_W \vert i \rangle \vert^2 \, 2\pi\,\delta(E_f-E_i)\,,
\end{equation}
where the sum runs over the neutrino helicites $h$, the integration is over the phase space volume $\Phi$; $\vert i\rangle$ and $\vert f\rangle$ are, respectively, the initial and final states consisting of leptons and nucleons, and $E_{i,f}$ are their energies~\cite{walecka}. At energy scales well below the masses of the $W^\pm$ and $Z^0$ bosons, the nuclear weak interaction Hamiltonian $\hat H_W$ can be written as a contact interaction between  the leptonic and nuclear current operators, 
\begin{equation}
\label{eq:weak_int_ham}
 \hat H_W = \frac{G}{\sqrt{2}}\int\mathrm{d}^3x \, j^\mathrm{lept}_\mu(\mathbf{x}) \, {j^\mu}(\mathbf{x})\,,
\end{equation}
where $G$ is the coupling constant. While the matrix elements of the leptonic operator $j^\mathrm{lept}_\mu$ are well approximated by free-space Dirac currents, the derivation of the nuclear operator $j^\mu$ and the calculation of its matrix element for nuclear states present challenges. The current operator $j^\mu$ and the nuclear wave functions have traditionally been obtained from phenomenological models with  hadronic degrees of freedom. Over the last few decades, $\chi$EFT has emerged as a successful theory that connects properties of nucleons and mesons to the underlying dynamics of quarks and gluons in a model-independent and systematically improvable way~\cite{Weinberg:1990rz,Weinberg:1991um,Epelbaum:1999dj,Entem:2003ft,Epelbaum:2008ga,Machleidt:2011zz,Carlsson:2015vda}. The nuclear wave functions are obtained from the $\chi$EFT interactions  arranged as a hierarchy of Feynman diagrams with interacting pion and nucleon (N) fields. The weak current operator $j^\mu$ is similarly expressed as couplings of the external sources to the $\pi$ and N fields and their interaction vertices within the same formalism and indeed shares several LECs with the strong-interaction Hamiltonian. $\chi$EFT thus provides a consistent theoretical framework in which both the interactions and the currents are organized in $Q/\Lambda_b$ expansions, where $Q$ is of the order of the pion mass $m_\pi$ and $\Lambda_b$ is the chiral symmetry-breaking scale which is roughly of the order of 1~GeV. We note, however, that a fully consistent treatment of interactions and current requires the use of the same regularization scheme, which is still under development~\cite{Krebs:2019uvm} and is beyond the scope of this work. 

\subsection{The neutrino dissociation cross section}

The differential cross section for the disintegration of the deuteron by an antineutrino or a neutrino of energy $\epsilon$, which follows from Eqs.~\eqref{eq:cross_sec_schematic} and \eqref{eq:weak_int_ham}, can be written in terms of the nuclear electroweak response functions $R_{\alpha\beta}$ as 

\begin{align}
\label{eq:diffcrosssec}
\frac{\mathrm{d}^2\sigma}{\mathrm{d}\Omega\,\mathrm{d}\omega}\bigg\vert_{\nu/\bar{\nu}}
=  \frac{G^2}{8\pi^2} \, & \frac{k^\prime}{\epsilon} \, F(Z,k^\prime) \, [v_{00} R_{00} + v_{zz} R_{zz} 
              - v_{0z} R_{0z}  \nonumber\\
              &  + v_{xx+yy} R_{xx+yy}
              \mp v_{xy} R_{xy}] \,.
\end{align}
The coupling constant $G$ is equal to the Fermi coupling $G_F$ for the neutral current (NC) process and to $G_F V_{ud}$, where $V_{ud}$ is the Cabibbo-mixing matrix element, for the charge-changing (CC) process; $k^\prime$~($\epsilon^\prime$) is the momentum~(energy) of the scattered lepton in the rest frame of the deuteron, and the function $F(Z,k^\prime)$, whose expression is given in Ref.~\cite{Feenberg:1950vj}, accounts for the distortion of the wave function of the final-state lepton due to the electric field of the nucleons.
The expressions for the lepton tensors $v_{\mu\nu}$, which can be obtained from Dirac algebra, are 
\begin{eqnarray}
\nonumber
 v_{00} & =& 2\epsilon\epsilon^\prime\left(1 + \frac{k^\prime}{\epsilon^\prime} \cos\theta\right)\,,\\
\nonumber
 v_{zz} & =& \frac{\omega^2}{q^2}\,(m_l^2+v_{00})+ 
           \frac{m_l^2}{q^2}\,[m_l^2+2\omega(\epsilon+\epsilon^\prime)+q^2] \,, \\
\nonumber
 v_{0z} & =& \frac{\omega}{q} \, (m_l^2+v_{00}) 
 + m_l^2 \, \frac{\epsilon+\epsilon^\prime}{q}\,, \\
\nonumber
 v_{xx+yy} & =& Q^2 + \frac{Q^2}{2q^2} (m_l^2+v_{00}) 
 - \frac{m_l^2}{q^2} \left[\frac{m_l^2}{2} + \omega(\epsilon+\epsilon^\prime)\right]\,, \\
 v_{xy} & = &Q^2 \, \frac{\epsilon+\epsilon^\prime}{q}-m_l^2\,\frac{\omega}{q}\,,
\end{eqnarray}
where the final-state lepton mass $m_l$ is equal to the electron mass for the CC process and zero for the NC process. The energy transfer is
\begin{equation}
 \omega = \epsilon - \epsilon^\prime\,,
\end{equation}
and the  magnitude of the three-momentum transfer is 
\begin{equation}
q=(\epsilon^2+{k^\prime}^2-2\,\epsilon \, k^\prime \, \cos\theta)^{1/2}\,,
\end{equation}
where $\theta$ is the scattering angle. The squared four-momentum transfer $Q^2$ is defined as $ Q^2 = -q^\mu q_\mu = q^2 - \omega^2>0$. For a monochromatic $\bar\nu/\nu$ beam of incident energy $\epsilon$, the differential cross section in Eq.~\eqref{eq:diffcrosssec} is, therefore, a function of only two kinematic variables: $\epsilon^\prime$ and $\theta$.

We choose the $z$ axis along the direction of $\mathbf{q}$ and the $zx$ plane along the plane of $\mathbf{q}$ and the   relative momentum $\mathbf{p}$ between the final-state nucleons. The magnitude of $\mathbf{p}$ is given up to corrections of ${O}\left(p^2q^2/m^4\right)$ by 
\begin{equation}
  (\omega+m_d)^2-q^2 = 4(p^2+m^2)\,,
\end{equation}
where $m_d$  and $m$ are the masses of the deuteron and nucleon, respectively. In case of the deuteron, the response functions $R_{\alpha\beta}$, which depend on $\omega$ and $q$, can be written as
\begin{align}
  R_{00} (\omega,q) = \frac{p^2}{24\pi^2} & \sum_{M_d}  \sum_{S^\prime S_z^\prime} \sum_{T^\prime} \int_{-1}^1\mathrm{d}x \nonumber\\
   & \qquad \frac{\vert \langle\psi_{\mathbf{p},S^\prime S_z^\prime,T^\prime T_z^\prime}\vert \rho\vert\psi_{d,M_d}\rangle \vert^2}{\big\vert \frac{p+xq/2}{E_+} + \frac{p-xq/2}{E_-} \big\vert}\,,
\end{align}
\begin{align}
  R_{zz}(\omega,q) = \frac{p^2}{24\pi^2} & \sum_{M_d} \sum_{S^\prime S_z^\prime} \sum_{T^\prime} \int_{-1}^1\mathrm{d}x \nonumber\\
         & \qquad \frac{\vert \langle\psi_{\mathbf{p},S^\prime S_z^\prime,T^\prime T_z^\prime}\vert j_0\vert\psi_{d,M_d}\rangle \vert^2}
         {\big\vert \frac{p+xq/2}{E_+} + \frac{p-xq/2}{E_-} \big\vert}\,,
\end{align}
\begin{align}
  & R_{0z}(\omega,q) = \frac{p^2}{24\pi^2} \sum_{M_d} \sum_{S^\prime S_z^\prime} \sum_{T^\prime} \int_{-1}^1 \mathrm{d}x  \nonumber\\
         & \quad \frac{2\,\Re \big\{ \langle\psi_{\mathbf{p},S^\prime S_z^\prime,T^\prime T_z^\prime}\vert \rho \vert\psi_{d,M_d}\rangle\langle\psi_{\mathbf{p},S^\prime S_z^\prime,T^\prime T_z^\prime}\vert j_0\vert\psi_{d,M_d}\rangle^\ast\big\}}{\big\vert \frac{p+xq/2}{E_+} + \frac{p-xq/2}{E_-} \big\vert}\,,
\end{align}
\begin{align}
  & R_{xx+yy}(\omega,q) = \frac{p^2}{24\pi^2} \sum_{M_d} \sum_{S^\prime S_z^\prime} \sum_{T^\prime} \int_{-1}^1\mathrm{d}x \nonumber\\
         & \, \frac{\vert \langle\psi_{\mathbf{p},S^\prime S_z^\prime,T^\prime T_z^\prime}\vert j_1\vert\psi_{d,M_d}\rangle \vert^2 + \vert \langle\psi_{\mathbf{p},S^\prime S_z^\prime,T^\prime T_z^\prime}\vert j_{-1}\vert\psi_{d,M_d}\rangle \vert^2}
         {\big\vert \frac{p+xq/2}{E_+} + \frac{p-xq/2}{E_-} \big\vert},
\end{align}
and
\begin{align}
  & R_{xy}(\omega,q) 
          = \frac{p^2}{24\pi^2} \sum_{M_d} \sum_{S^\prime S_z^\prime} \sum_{T^\prime} \int_{-1}^1\mathrm{d}x \nonumber\\
         &\frac{\vert \langle\psi_{\mathbf{p},S^\prime S_z^\prime,T^\prime T_z^\prime}\vert j_1\vert\psi_{d,M_d}\rangle \vert^2 - \vert \langle\psi_{\mathbf{p},S^\prime S_z^\prime,T^\prime T_z^\prime}\vert j_{-1}\vert\psi_{d,M_d}\rangle \vert^2}{\big\vert \frac{p+xq/2}{E_+} + \frac{p-xq/2}{E_-} \big\vert}\,.
\end{align}
Here the operator $\rho$ is the zeroth component of the four-vector weak current and $j_\lambda$ are the spherical components of the three-vector weak current operator ${\bf j}$. The integration variable $x$ is the cosine of the angle between ${\bf q}$ and ${\bf p}$. 
The initial nuclear state is the deuteron ground state, denoted here by $| \psi_{d,M_d}\rangle$,  where $M_d$ is the projection of the total angular momentum, while
the final nuclear state is denoted by $|\psi_{\mathbf{p},S^\prime S_z^\prime,T^\prime T_z^\prime}\rangle $, where $T^\prime$, $T_z^\prime$, $S^\prime$,  $S_z^\prime$ are, respectively, the total isospin, isospin projection, total spin and spin projection of the scattering two-body state. Finally, $E_\pm = \sqrt{(\mathbf{q}/2 \pm \mathbf{p})^2+m^2}$ are their energies in the rest frame of the deuteron. 
 
At this point, it is convenient to perform a multipole decomposition of the operators $\rho$ and $j_\lambda$. This can be used for the deuteron calculations presented in this paper, but it is also applicable to computations in heavier nuclei, where one typically uses a spherical basis.
Within this formalism, the matrix elements  of the charge/current operators can be expanded in terms of  reduced matrix elements  of spherical tensor operators, i.e., the multipoles of $\rho$ and $j_\lambda$, as
\begin{align}
\label{eq:j_t_mat_elem}
 & \langle \psi_{\mathbf{p},S^\prime S_z^\prime,T^\prime T_z^\prime} \vert \rho \vert \psi_{d,M_d} \rangle 
= (4\pi)^{3/2}\sqrt{2} \sum_{\Lambda J^\prime L_d L L^\prime} i^{\Lambda-L} \nonumber\\ 
& \qquad\qquad\quad(-1)^{1+S^\prime+\Lambda-L} \, [\Lambda]\, [J^\prime] \, {Y_{L}^{M_d-S_z^\prime}}(\hat p) \nonumber\\
 & \qquad\qquad \threejm{L}{S^\prime}{J^\prime}{L_z}{S_z^\prime}{-M_d} 
  \threejm{1}{\Lambda}{J^\prime}{M_d}{0}{-M_d} \nonumber\\
 & \qquad\qquad _{L}\langle p;(L^\prime S^\prime)J^\prime;T^\prime T_z^\prime \vert\vert \mathcal{C}_\Lambda \vert\vert (L_d1)1;00 \rangle
 \end{align}
and 
\begin{align}
\label{eq:j_lambda_mat_elem}
 & \langle \psi_{\mathbf{p},S^\prime S_z^\prime,T^\prime T_z^\prime} \vert j_\lambda \vert \psi_{d,M_d} \rangle 
 = -(4\pi)^{3/2} \sum_{\Lambda J^\prime L_d L L^\prime}
 i^{\Lambda-L}  \nonumber\\ 
 &  \qquad\qquad (-1)^{1+S^\prime+\Lambda-L} \, [\Lambda]\, [J^\prime] \, {Y_{L}^{M_d+\lambda-S_z^\prime}}(\hat p) \nonumber\\
 & \qquad\,\,\threejm{L}{S^\prime}{J^\prime}{L_z}{S_z^\prime}{-M_d-\lambda} \threejm{1}{\Lambda}{J^\prime}{M_d}{\lambda}{-M_d-\lambda} \nonumber\\
 & \qquad \big[ \sqrt{2}\,\,_{L}\langle p;(L^\prime S^\prime)J^\prime;T^\prime T_z^\prime \vert\vert\mathcal{L}_\Lambda\vert\vert (L_d1)1;00 \rangle\,\delta_{0\lambda}+\nonumber\\
 & \qquad\, \,_{L}\langle p;(L^\prime S^\prime)J^\prime;T^\prime T_z^\prime \vert\vert \lambda\mathcal{M}_\Lambda+\mathcal{E}_\Lambda \vert\vert (L_d1)1;00 \rangle\,\delta_{\pm1\lambda}\big]\,.
 \end{align}
 Here we have used the three-$j$ symbol~\cite{VMK}; $Y_L^\mu$ is a spherical harmonics of generic multipolarity $L$ and projection $\mu$,  while $[\Lambda]$ denotes $\sqrt{2\Lambda+1}$.

In these expressions,  $\mathcal{C}_\Lambda^M$, $\mathcal{L}_\Lambda^M$, $\mathcal{E}_\Lambda^M$ and $\mathcal{M}_\Lambda^M$ are, respectively, the Coulomb, longitudinal, transverse electric, and transverse magnetic multipole operators~\cite{walecka} defined in terms of $\rho$ and ${\bf j}$ as 

\begin{align}
 \mathcal{C}_\Lambda^M &= \frac{(-i)^\Lambda}{4\pi} \int \mathrm{d}\Omega_{\hat q} \, Y_\Lambda^M(\hat q) \, \rho\,,\\
  \mathcal{L}_\Lambda^M &= i\left(\frac{\sqrt{\Lambda}}{[\Lambda]}\mathcal{D}_{\Lambda,\Lambda-1}^M
 +\frac{\sqrt{\Lambda+1}}{[\Lambda]}\mathcal{D}_{\Lambda,\Lambda+1}^M\right)\,,\\
  \mathcal{E}_\Lambda^M &= i\left(\frac{\sqrt{\Lambda+1}}{[\Lambda]}\mathcal{D}_{\Lambda,\Lambda-1}^M
 -\frac{\sqrt{\Lambda}}{[\Lambda]}\mathcal{D}_{\Lambda,\Lambda+1}^M\right)\,, \\
 \mathcal{L}_\Lambda^M &= i\left(\frac{\sqrt{\Lambda}}{[\Lambda]}\mathcal{D}_{\Lambda,\Lambda-1}^M
 +\frac{\sqrt{\Lambda+1}}{[\Lambda]}\mathcal{D}_{\Lambda,\Lambda+1}^M\right)\,,\\
 \mathcal{M}_\Lambda^M &= \mathcal{D}_{\Lambda,\Lambda}^M\,,
 \end{align}
 where
\begin{equation}
 \mathcal{D}_{\Lambda,K}^M = \frac{(-i)^K}{4\pi} \int \mathrm{d}\Omega_{\hat q} \, \mathcal{Y}_{\Lambda (K1)}^M(\hat q) \cdot \mathbf{j}\,.
\end{equation}
The deuteron ground state $\vert \psi_{d,M_d} \rangle$ can be written in coordinate representation as an expansion in partial waves:
\begin{align}
 \langle \mathbf{r} \vert \psi_{d,M_d} \rangle
 &=\sum_{L_d=0,2}\langle \mathbf{r} \vert(L_d1)1 M_d;00\rangle\nonumber\\
 &= \sum_{L_d=0,2} \frac{u_{L_d}(r)}{r} \, \mathcal{Y}_{1(L_d1)}^{M_d}(\hat r) \, \vert T=0, T_z=0\rangle\,,
\end{align}
where $ \mathcal{Y}_{\Lambda (K1)}^M(\hat q)$ are vector spherical harmonics~\cite{VMK} and $u_{0,2}(r)$ are the deuteron radial wave functions.
The NN scattering state $\vert p;(L^\prime S^\prime)J^\prime;T^\prime T_z^\prime \rangle$ is similarly given by
\begin{align}
  & \langle \psi_{\mathbf{p},S^\prime S_z^\prime,T^\prime T_z^\prime}  \vert \mathbf{r}\rangle 
  = 4\pi\sqrt{2} \sum_{J^\prime J_z^\prime L^\prime L L_z} i^{-L}\,{\mathcal{Y}_{J^\prime(L^\prime S^\prime)}^{J_z^\prime}}^\ast (\hat r) \nonumber\\
 & \qquad\qquad {Y_{L}^{L_z}}(\hat p)\, \langle L L_z; S^\prime S_z^\prime \vert (L S^\prime) J^\prime J_z^\prime \rangle \, {z^{J^\prime S^\prime T^\prime}_{L^\prime L}}^\ast(pr)\nonumber\\
 & \qquad \equiv 4\pi\sqrt{2} \sum_{J^\prime J_z^\prime L^\prime L L_z} i^{-L} \langle L L_z; S^\prime S_z^\prime \vert (L S^\prime) J^\prime J_z^\prime \rangle\nonumber\\
 & \qquad\qquad  {Y_{L}^{L_z}}(\hat p)\, \,_{L}\langle p;(L^\prime S^\prime)J^\prime J_z^\prime;T^\prime T_z^\prime\vert\mathbf{r}\rangle\,,
\end{align}
where $\langle L L_z; S^\prime S_z^\prime \vert (L S^\prime) J^\prime J_z^\prime \rangle $ is a  Clebsch-Gordan coefficient~\cite{VMK}.
The radial wave functions of the scattering state, ${z^{J^\prime S^\prime T^\prime}_{L^\prime L}}(pr)$, have the asymptotic form 
\begin{align}
\label{eq:nocoul_asymp}
   z^{J^\prime S^\prime T^\prime}_{L^\prime L}(pr) 
  \rightarrow \frac{1}{2}\big[& \delta_{L^\prime L}\,h^{(2)}_L(\eta;pr) \nonumber\\
  & + h^{(1)}_{L^\prime}(\eta;pr)\,S^{J^\prime S^\prime T^\prime}_{L^\prime L}(p,p)\big]\,, 
\end{align}
where $S^{J^\prime S^\prime T^\prime}_{L^\prime L}(p,p)$ is the scattering matrix and $h^{(1,2)}_{L}(\eta;pr)$ are outgoing and 
incoming Coulomb wave functions at Sommerfeld parameter $\eta$. For the $nn$ and $pn$ systems, $\eta=0$ and the functions $h^{(1,2)}_{L}(\eta=0;pr)$, therefore, reduce to spherical Hankel functions. 
The radial wave functions $u_{0,2}(r)$ and $z^{J^\prime S^\prime T^\prime}_{L^\prime L}(pr)$ are obtained by solving the partial wave Lippmann-Schwinger equation as outlined in Refs.~\cite{shen,Schiavilla:2004wn,Carlson:2001ma}. The reduced multipole matrix elements in Eqs.~\eqref{eq:j_t_mat_elem} and \eqref{eq:j_lambda_mat_elem} are numerically evaluated by truncating the summation over multipolarity $\Lambda$ and are then used to obtain the nuclear electroweak response functions $R_{\alpha\beta}(\omega,q)$ for a discrete mesh of $\omega$ and $q$. The number of multipoles required depends on the value of $q$. We find that converged results are obtained for the range of kinematics considered in this work with $\Lambda$ up to 10. 

The total cross section $\sigma(\epsilon)$ can be obtained by integrating Eq.~\eqref{eq:diffcrosssec} over $\theta$ and $\epsilon^\prime$. 
The limits on the $\epsilon^\prime$ integrals are set by the 
kinematical constraints $m_l\leq\epsilon^\prime\leq\epsilon^\prime_+$ for $0\leq\theta\leq\pi/2$ and $m_l\leq\epsilon^\prime\leq\epsilon^\prime_-$ for $\pi/2\leq\theta\leq\pi$. 
Here the upper limits $\epsilon^\prime_\pm$ are given by 
\begin{equation}
 \epsilon^\prime_\pm = \frac{\bar{\epsilon}\,\pm\,
 [\bar{\epsilon}^2-(1-\beta^2\cos^2\theta)(\bar{\epsilon}^2+m_l^2\beta^2\cos^2\theta)]^{1/2}
 }{1-\beta^2\cos^2\theta}\,,
\end{equation}
where
\begin{equation}
 \beta=\frac{\epsilon}{\epsilon+m_d}\,,
\end{equation}
and
\begin{equation}
 \bar{\epsilon}=\frac{m_d(\epsilon-\epsilon_{th})+m_l(m_l+2m)}{\epsilon+m_d}\,.
\end{equation}
The threshold energy of the incident neutrino is 
\begin{equation}
 \epsilon_{th}=\frac{(m_l+2m)^2-m_d^2}{2m_d}\,,
\end{equation}
where $m$ is $(m_p+m_n)/2$ for NC processes, $m_p$ for CC $\nu$ scattering, and $m_n$ for CC $\bar{\nu}$ scattering. 

\subsection{The current operators in $\chi$EFT}

The electroweak current operators were first derived within the context of $\chi$EFT in Refs.~\cite{Park:1995pn,park_axial,phillips}. More general and complete derivations were later performed using the unitary transformation method~\cite{Kolling:2009iq,Kolling:2011mt,Krebs:2016rqz,Krebs:2019aka,Krebs:2019uvm} and in many-body perturbation theory~\cite{Pastore:2008ui,Pastore:2009is,Pastore:2011ip,Baroni:2015uza}. The operators we use in this work are consistent with both of these sets of studies because the differences that exist between them do not appear up to the chiral order at which we work. 
As in Ref.~\cite{Phillips:2016mov}, we count the inverse nucleon mass~($1/m$) factors that arise from Gordon decomposition of the Dirac current as one chiral order and relativistic $1/m^2$ corrections as four chiral orders. This is different from both Refs.~\cite{Kolling:2009iq,Kolling:2011mt,Krebs:2016rqz,Krebs:2019aka,Krebs:2019uvm} that count $m$ as ${O}(\Lambda_b^2/Q)$ and Refs.~\cite{Pastore:2008ui,Pastore:2009is,Pastore:2011ip,Baroni:2015uza} that count it as ${O}(\Lambda_b)$, but does not lead to inconsistencies with the power counting of operators in the strong-interaction Hamiltonian. 

We now provide a brief overview of the forms of the current operators that we will implement. The neutral weak current is given by 
$j_\mathrm{NC}^\mu = -2\sin^2\theta_W \, j_{\gamma,S}^\mu + \left(1-2\sin^2\theta_W\right) j_{\gamma,z}^\mu + j_z^{\mu5}$,
where $\theta_W$ is the Weinberg angle, $j_{\gamma,S}^\mu$ and $j_{\gamma,z}^\mu$ are the isoscalar and isovector electromagnetic currents, 
and $j_z^{\mu5}$ is the weak axial current, whereas the charge-changing weak current operator, $j_\mathrm{CC}^{\mu}$, can be written as the sum of the 
vector and the axial vector pieces, $j_\pm^{\mu}+j_\pm^{\mu5}$. Each of these terms can be expressed as a sum of one-body (1B) and two-body (2B) operators that act on nucleonic degrees of freedom as
\begin{equation}
 j^\mu = \sum_{n} j^\mu(n) + \sum_{m<n} j^\mu(mn)\,,
\end{equation}
where the sums run over the nucleons.

We consider all electroweak operators  at orders $(Q/\Lambda_b)^{-3,-2,-1,0}$ in the $\chi$EFT power counting. The leading 1B vector charge operator occurs at $(Q/\Lambda_b)^{-3}$. Its expression is 
\begin{equation}
 j_{\gamma,S/z}^0 (n) = G_E^{S/V}(Q^2) \dfrac{1}{\sqrt{1+\frac{Q^2}{4m^2}}} e^{i\mathbf{q}\cdot\mathbf{r}_n} \, \tau_n^{S/V}\,,
\end{equation}
where $\mathbf{r}_n$ is the position of the $n$-th nucleon. The isoscalar isospin operator $\tau_n^S$ is one-half times the the identity operator whereas the isovector isospin operator $\tau_n^V$ is $\tau_{n,z}/2$. 

The isoscalar and isovector electric form factors can be written in terms of the proton and neutron electric form factors as  $G_E^{S/V}= G_E^p \pm G_E^n$. At least up to the chiral order at which we work, the nucleon structure corrections that occur for the 1B parts of the current operator calculated between two-body states are exactly the same as those for free protons and neutrons. These nucleon-structure corrections have been derived in chiral effective field theory~\cite{Kubis:2000zd}. However, several orders of calculations are needed to obtain converged results. It has therefore become a common practice to use phenomenological form factors to represent the sum of the nucleon structure diagrams, which makes the calculations of nuclear systems less sensitive to inaccuracies in the single-nucleon sector~\cite{Phillips:2016mov}. We use the dipole parametrization of the electromagnetic form factors with a vector mass factor of 833~MeV as in Refs.~\cite{shen,Baroni:2017gtk}. 

The 1B vector current operator first contributes at ${O}(Q/\Lambda_b)^{-2}$. It consists of the so-called convection and spin-magnetization currents,
\begin{align}
 \mathbf{j}_{\gamma,S/z}(n) =  \bigg( & G_E^{S/V}(Q^2)\frac{\mathbf{\bar{p}}_n}{m} 
   - i \, G_M^{S/V}(Q^2)\frac{\mathbf{q}\times\bm{\sigma}_n}{2m}\bigg) \nonumber\\
   & \,\,e^{i\mathbf{q}\cdot\mathbf{r}_n} \, \tau_n^{S/V} \,,
\end{align}
where $G_M^{S/V}= G_M^p \pm G_M^n$ are the isoscalar and isovector magnetic form factors.  
The momentum of the $n$th nucleon, $\mathbf{\bar{p}}_n=(\mathbf{p}_n^\prime+\mathbf{p}_n)/2 = \mathbf{p}_n+\mathbf{q}/2$ is the average 
of its initial and final momenta. 

The 1B axial current is given at ${O}(Q/\Lambda_b)^{-3}$ by 
\begin{equation}
 \mathbf{j}_z^{5}(n) = - G_A(Q^2) \, \bm{\sigma}_n \, e^{i\mathbf{q}\cdot\mathbf{r}_n} \, \tau_n^V\,,
\end{equation}
and the 1B axial charge at ${O}(Q/\Lambda_b)^{-2}$ by
\begin{equation}
 j_z^{05}(n) = - G_A(Q^2) \, \bm{\sigma}_n \cdot \frac{\mathbf{\bar{p}}_n}{m} \, e^{i\mathbf{q}\cdot\mathbf{r}_n} \, \tau_n^V\,.
\end{equation}
Here $\bm{\sigma}_n$ is the Pauli operator acting on the nucleon spin and $G_A(Q^2)$ is the axial form factor. It was recently claimed that the dipole parametrization of $G_A(Q^2)$ yields large systematic deviations from the $z$ expansion~\cite{Bhattacharya:2011ah}. Therefore, in addition to a dipole parametrization with axial mass $M_A=1$~GeV, we also use a model-independent expansion of the axial form factor,
$G_A(Q^2) = g_A\left[1-\langle r_A^2\rangle\,Q^2/6\right]+{O}(Q^4)$, where $g_A$ is the axial coupling constant and $\langle r_A^2\rangle$ is the mean-square axial radius of the nucleon. It is to be noted that the ${O}(Q^4)$ corrections enter at an order beyond the maximum $\chi$EFT order we consider for our electroweak operators.

The charge-changing operator $j_\mathrm{CC}^{\mu}(n)=j_\pm^{\mu}(n)+j_\pm^{\mu5}(n)$ can be obtained from $j_{\gamma,z}^\mu(n)+j_z^{\mu5}(n)$ by 
the substitution 
\begin{equation}
  \tau_n^V=\frac{\tau_{n,z}}{2} \rightarrow \frac{\tau_{n,x} \pm i\,\tau_{n,y}}{2}=\tau_{n,\pm}\,,
\end{equation}
along with the inclusion of induced pseudoscalar contributions, for which we use the expression given in terms of the axial form factor,  
 \begin{equation}
  j_\pm^{\mu5} (n;\mathrm{PS}) = G_A(Q^2)\frac{q^\mu\,\bm{\sigma}_n\cdot\mathbf{q}}{m_\pi^2+Q^2} \, e^{i\mathbf{q}\cdot\mathbf{r}_n} \, \tau_{n,\pm}\,,
 \end{equation}
using the parametrization obtained from chiral Ward identity~\cite{Bernard:1994wn}.

\begin{table*}[t]
\centering
\begin{tabular*}{\textwidth}{@{\extracolsep{\fill}}|l|cccc|cccc|cccc|cccc|}
\toprule
&\multicolumn{4}{c|}{$\nu$, NC}&\multicolumn{4}{c|}{$\nu$, CC}&\multicolumn{4}{c|}{$\bar\nu$, NC}&\multicolumn{4}{c|}{$\bar\nu$, CC}\\
\hline
~~~~~~~~~~~~$\epsilon$ [MeV]&10&50&100&150&10&50&100&150&10&50&100&150&10&50&100&150\\
~~~~~~~~~~~~~~~~~x&16&15&14&14&16&14&14&13&16&15&14&14&16&15&14&14\\
 \hline
 EM500/1B/$(Q/\Lambda_b)^{-3}$ &1.04&5.11&2.08&4.27&2.50&1.05&4.38&0.92&1.04&5.11&2.08&4.27&1.25&9.28&4.02&8.53\\
  EM500/1B/$(Q/\Lambda_b)^{-2}$ &1.07&5.80&2.61&5.81&2.62&1.32&6.46&1.53&1.02&4.47&1.61&3.01&1.20&7.33&2.61&4.81\\
  EM500/1B+2B/$(Q/\Lambda_b)^{-1}$ &1.07&5.82&2.62&5.85&2.62&1.33&6.53&1.55&1.02&4.46&1.60&3.00&1.20&7.30&2.60&4.79\\
    EM500/1B+2B/$(Q/\Lambda_b)^{0}$&1.10&6.01&2.71&6.07&2.70&1.36&6.69&1.59&1.05&4.62& 1.67&3.17&1.23&7.57 &2.71&5.07\\
\hline
 EM500/1B+2B~(Ref.~\cite{Baroni:2017gtk}) &1.12&6.03&2.74&6.18&2.73&1.39&6.85&1.65&1.07&4.63&1.68&3.21&1.27&7.52&2.68&4.98\\
 \hline
 AV18/1B~(Ref.~\cite{shen})
 &1.08&5.75&2.58&5.72&2.63&1.31&6.42&1.51&1.03&4.45&1.60&3.00&1.22&7.26&2.57&4.69\\
AV18/1B+2B~(Ref.~\cite{shen}) &1.10&5.89&2.66&5.94&2.68&1.35&6.63&1.57&1.05&4.55&1.64&3.08&1.24&7.40&2.61&4.75\\
\botrule
  \end{tabular*}
\caption{Inclusive $\bar\nu/\nu$-$d$ cross sections. The values are for energy $\epsilon$~MeV in units of $10^{-\mathrm{x}}\,\mathrm{fm}^2$ with $\epsilon$ and $\mathrm{x}$ values given in the corresponding columns. EM500/1B/$(Q/\Lambda_b)^m$ [EM500/1B+2B/$(Q/\Lambda_b)^m$] stands for a calculation that employs the EM500 interaction to generate the wave functions and includes all 1B [1B and 2B] currents up to the order $(Q/\Lambda_b)^m$. The AV18/1B calculation of Ref.~\cite{shen} uses the same current operators as the EM500/1B/$(Q/\Lambda_b)^{-2}$ calculation. The EM500/1B+2B calculation of Ref.~\cite{Baroni:2017gtk} also includes currents up to $(Q/\Lambda_b)^1$. \label{tab:crosssec}}
\end{table*}

The 2B vector current operator is purely isovector up to the order we consider. The one-pion-exchange operators enter at ${O}(Q/\Lambda_b)^{-1}$. They are given by the sum of the so-called seagull and pion-in-flight terms, which can be written in momentum space as 
\begin{align}
\mathbf{j}_{\gamma,z}(mn) = -i \, &  \frac{g_A^2}{4f_\pi^2}  
\bigg( \bm{\sigma}_m-\mathbf{k}_m\frac{\bm{\sigma}_m\cdot\mathbf{k}_m}{m_\pi^2+k_m^2}\bigg)\nonumber\\ & \,\frac{\bm{\sigma}_n\cdot\mathbf{k}_n}{m_\pi^2+k_n^2}
 \left(\bm{\tau}_m\times\bm{\tau}_n\right)_z + (m\leftrightarrow n)\,,
\end{align}
where $\mathbf{k}_n=\mathbf{p}^\prime_n-\mathbf{p}_n$, $f_\pi$ is the pion-decay constant and $g_A$ is the axial coupling constant. The 2B axial charge, 
\begin{align}
 j_z^{05} (mn) = -i\,\frac{g_A}{4f_\pi^2} \frac{\bm{\sigma}_m\cdot\mathbf{k}_m}{m_\pi^2+k_m^2} \left(\bm{\tau}_m\times\bm{\tau}_n\right)_z + (m \leftrightarrow n) \,,
 \end{align}
 enters at the same order. 
At the third chiral order, i.e., at ${O}(Q/\Lambda_b)^0$, we have the 2B axial current. These include the one-pion exchange operators, some of which contain the dimensionless $\pi$N couplings $\hat c_{1,3,4}$, and the 2B contact current with LECs $\hat d_{1,2}$. These can be combined into the expression 
\begin{align}
 \mathbf{j}_z^{5} (mn)= \frac{g_A}{2mf_\pi^2}  &\frac{\bm{\sigma}_n\cdot\mathbf{k}_n}{m_\pi^2+k_n^2}\nonumber\\
  &\bigg[
  \frac{i}{2} \, \mathbf{\bar p}_m \left(\bm{\tau}_m\times\bm{\tau}_n\right)_z + 4\hat c_3 \, \mathbf{k}_n \frac{\tau_{n,z}}{2} \nonumber\\
 & +\left(\hat c_4+\frac{1}{4}\right)\bm{\sigma}_m\times\mathbf{k}_n\left(\bm{\tau}_m\times\bm{\tau}_n\right)_z \nonumber\\
 & +\frac{\mu_V}{4}\bm{\sigma}_m\times\mathbf{q}\left(\bm{\tau}_m\times\bm{\tau}_n\right)_z\bigg] \nonumber\\
 &+2 \hat d_1 (\bm{\sigma}_m\frac{\tau_{m,z}}{2}+\bm{\sigma}_n\frac{\tau_{n,z}}{2}) \nonumber\\
 & + \hat d_2\,\bm{\sigma}_m\times\bm{\sigma}_n\left(\bm{\tau}_m\times\bm{\tau}_n\right)_z\nonumber\\
 & +(m\leftrightarrow n)\,.
\end{align} 
The forms of the contact operators are such that their matrix elements can only contain the linear combination $\hat d_1 + 2\hat d_2+\hat c_3/3+2\hat c_4/3+1/6$ for antisymmetric wave functions. This combination is conventionally referred to as $\hat d_R$. It is related to the LEC $c_D$~\cite{Gazit:2008ma}, which features in the leading three-nucleon interaction along with the LEC $c_E$ and $\hat c_{1,3,4}$, by 
\begin{equation}
 \hat d_R = -\textstyle{\frac{m }{4g_A\Lambda_b}}c_D
+\frac{1}{3}\hat c_3 + \frac{2}{3}\hat c_4 +\frac{1}{6}\,.
\end{equation} 
To date, two-nucleon weak processes have not been measured with 
sufficient precision to allow an extraction of $\hat d_R$. There is an ongoing effort to measure the rate of muon capture on the deuteron~\cite{Andreev:2010wd}, which might address this issue~\cite{Acharya:2018qzk}. In this work, we use the values of $\hat d_R$ obtained by following two different approaches: (i) Calculations that employ the NN interactions of Refs.~\cite{Entem:2003ft,Machleidt:2011zz} use the value obtained by performing a fit of the counterterms $c_D$ and $c_E$ in the leading 3N potential~\cite{Machleidt:2011zz} to experimental values of binding energies of $^3$H and $^3$He  as well as the comparative  $\beta$-decay half-life of $^3$H with predetermined $\pi$N and NN couplings~\cite{Gazit:2008ma}. (ii) The NNLO$_\mathrm{sim}$ calculations fix  $\hat d_R$ by performing a simultaneous fit of all of the LECs up to the third $\chi$EFT order to $\pi$N and selected NN scattering data, the binding energies and charge radii of $^{2,3}$H and $^3$He, the quadrupole moment of $^2$H, as well as the $\beta$-decay half-life of $^3$H~\cite{Carlsson:2015vda,Andreas}. 

Finally, the 2B charge-changing weak current (CC) operator, 
$j_\mathrm{CC}^{\mu}(mn)=j_\pm^{\mu}(mn)+j_\pm^{\mu5}(mn)$, can be obtained from $j_{\gamma,z}^\mu(mn)+j_z^{\mu5}(mn)$ by 
the substitution 
\begin{align}
\frac{\tau_{n,z}}{2} & \rightarrow \tau_{n,\pm}\,, \nonumber\\
  \left(\bm{\tau}_m\times\bm{\tau}_n\right)_z & \rightarrow \left(\bm{\tau}_m\times\bm{\tau}_n\right)_x \pm i\, \left(\bm{\tau}_m\times\bm{\tau}_n\right)_y \,,
\end{align}
along with the addition of the pion-pole contribution,
$q^\mu\left[q_\nu j_\pm^{\nu5}(mn)+j_\pm^5(mn;\mathrm{PS})\right]/(m_\pi^2+Q^2) \,,$ 
 where 
\begin{equation}
 j_\pm^5(mn;\mathrm{PS}) = \frac{4g_Am_\pi^2}{mf_\pi^2}\,\hat c_1 \frac{\bm{\sigma}_m\cdot\mathbf{k}_m}{m_\pi^2+k_m^2} \tau_{m,\pm}+(m \leftrightarrow n)\,.
\end{equation} 

Coordinate space expressions are obtained by Fourier transformations using the Gaussian regulators of the form $\exp[-1/2 \, (k_{1,2}/\Lambda)^2]$. While these are different from the regulators used in the interactions~\cite{Entem:2003ft,Machleidt:2011zz,Carlsson:2015vda} which are Gaussian functions of the nucleon momenta, this regularization is common in the literature and is consistent with the one used in the currents for the extraction of $c_D$ from tritium $\beta$ decay.

\section{Results}
\label{sec:results}

\subsection{Benchmark with previous work}

We first benchmark our results with previous works. To this end, 
we use wave functions obtained from the nonlocal $\chi$EFT interaction of Refs.~\cite{Entem:2003ft,Machleidt:2011zz} (referred to as ``EM500'' hereafter). This interaction is calculated up to the fourth chiral order with a regulator cutoff of 500~MeV and reproduces the NN scattering data up to 290~MeV laboratory-frame energy with very high precision. Fixing the potential to a high chiral order facilitates the comparison with Refs.~\cite{Baroni:2017gtk} and \cite{shen} and helps one to assess of the size of the contributions of the various terms in the current operator.

In Table~\ref{tab:crosssec}, we show the CC- and NC-induced inclusive $\bar\nu/\nu$-$d$ cross sections  obtained using the EM500 interaction and current operators of various $\chi$EFT orders. The EM500 interactions contain all effects that are suppressed by factors of up to $(Q/\Lambda_b)^{4}$ compared to the leading order $\chi$EFT Hamiltonian. With wave functions obtained by solving the partial wave Lippmann-Schwinger equations for this interaction, we vary the order of the weak current operator at $(Q/\Lambda_b)^{-3,-2,-1,0}$  to study the order-by-order convergence of the current in the $\bar\nu/\nu$-$d$ cross sections. 
With increasing energy, the 1B Fermi and Gamow-Teller operators, which contribute at the leading $(Q/\Lambda_b)^{-3}$ order,  underpredict (overpredict) the $\nu$-$d$ ($\bar\nu$-$d$) cross sections compared to values obtained with operators up to $(Q/\Lambda_b)^{0}$ order. The contributions of the 1B convection and spin-magnetization currents, which enter at order $(Q/\Lambda_b)^{-2}$, amount to about 30\% in the $\epsilon\approx100$~MeV region.
The pion-exchange 2B contributions to the vector current and axial charge operators, which formally enter at order $(Q/\Lambda_b)^{-1}$, are smaller than the axial 2B current contributions at $(Q/\Lambda_b)^0$. While this is contrary to expectations from $\chi$EFT power counting, a similar convergence pattern was also found by Ref.~\cite{Baroni:2017gtk}. Overall, the inclusion of 2B currents increases the cross section in all of the four reaction channels by about 3-4\% at $\epsilon\approx100$~MeV, which is consistent with the results of Ref.~\cite{Baroni:2017gtk}.

 Agreement is seen between our 1B results  and those of Ref.~\cite{shen}. The slight  difference of about 1\% or less is due to the AV18~\cite{Wiringa:1994wb} wave functions used by Ref.~\cite{shen}, since the $\chi$EFT 1B operators used in this work are the same as the phenomenological operators employed in that study. We  agree also within approximately 1\% with Ref.~\cite{Baroni:2017gtk}, which uses the same interactions for the wave functions but also includes the $(Q/\Lambda_b)^1$ current operators not considered in this work.

\subsection{Uncertainty estimates}

We now estimate, for the first time on this observable,  the uncertainty from the potential by using the NNLO$_\mathrm{sim}$ family of 42 interactions calculated up to the third chiral order~\cite{Carlsson:2015vda,Andreas}. These have been fitted at seven different values of the regulator cutoff $\Lambda$ in the 450-600~MeV interval to six different $T_{\rm lab}$ ranges in the NN scattering database. The LECs in this family of interactions were fitted {\it simultaneously} to $\pi$N and selected NN scattering data, the energies and charge radii of $^{2,3}$H and $^{3}$He, the quadrupole moment of $^{2}$H, as well as the $\beta$-decay width of $^{3}$H. All of these interactions have the correct long-range properties, and the differences between them provide a conservative estimate of the uncertainty due to the short-distance model ambiguity of $\chi$EFT.

\begin{figure}[htbp]
\begin{center}
\includegraphics[width=0.8\columnwidth,clip=true]{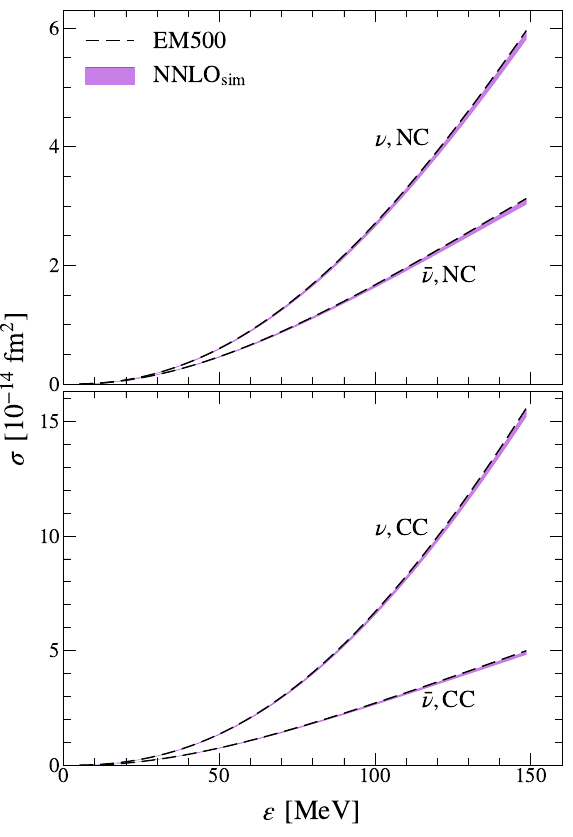} 
\end{center}
\caption{\protect (Color online) The NC and CC $\bar\nu/\nu$-$d$ inclusive cross sections with the EM500 (black, dashed) and NNLO$_\mathrm{sim}$ (light  band) interactions.}
\label{fig:nnlosim}
\end{figure}

 In Fig.~\ref{fig:nnlosim} we show, along with the EM500 curves, the cross sections calculated using the NNLO$_\mathrm{sim}$ interactions as bands.
 The widths of the bands are estimates of the uncertainties due to the sensitivity to the $\chi$EFT cutoff and variations in the pool of fit data used to constrain the LECs, including $\hat c_{1,3,4}$ and $\hat d_R$ in the currents.
These widths grow with $\epsilon$ and amount to about 3\% at $\epsilon\approx100$~MeV for all of the four processes. They are thus similar in size to the effect of 2B currents. The interactions and currents in the NNLO$_\mathrm{sim}$ results are of the same chiral order, i.e., both of them include all corrections that are suppressed by factors of up to $(Q/\Lambda_b)^3$ compared to the leading order. Based on the observed convergence of the cross sections in Table~\ref{tab:crosssec}, and on the results of Ref.~\cite{Baroni:2017gtk} for higher-order current contributions, 
we anticipate the size of neglected terms in the chiral expansion of the weak current operator to be 1\% at $\epsilon\approx100$~MeV. This is smaller than the NNLO$_\mathrm{sim}$ uncertainties, which are---in principle as well as in practice--- similar in size to the $(Q/\Lambda_b)^0$ current contributions which we have included in our calculations. We therefore assign a conservative estimate of 3\% to the nuclear structure uncertainties in the cross section at 100 MeV $\bar\nu$/$\nu$ energy.
\begin{figure}[ht]
\begin{center}
\includegraphics[width=0.8\columnwidth,clip=true]{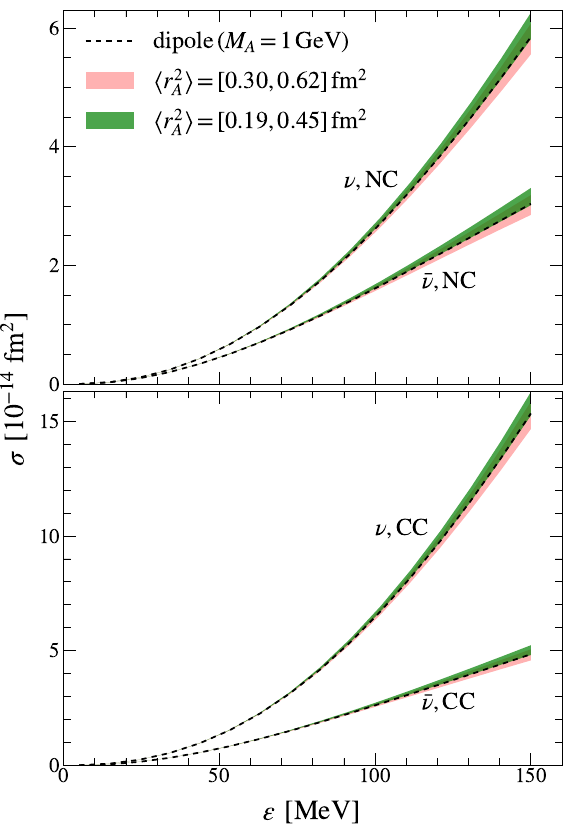} 
\end{center}
\caption{\protect (Color online) The 1B NC and CC $\bar\nu/\nu$-$d$ inclusive cross sections with dipole form factor for $M_A=1$~GeV (dotted) along with uncertainty bands from variation of $\langle r_A^2\rangle$ within the uncertainties of Ref.~\cite{Hill:2017wgb} (light band) and over the range of lattice QCD values (dark band).}
\label{fig:axial_ff}
\end{figure}
We now turn to the question of the sensitivity of these results to the single-nucleon axial form factor. Ref.~\cite{Meyer:2016oeg} analyzed the world data for $\nu d$ scattering by employing the calculations of Refs.~\cite{shen,Singh:1971md} to obtain $\langle r_A^2 \rangle=0.46\pm0.22~\mathrm{fm}^2$. Combining this with a reanalysis of the muon-proton capture data, Ref.~\cite{Hill:2017wgb} constrained the mean-squared axial radius to $0.46\pm0.16~\mathrm{fm}^2$. The nucleon axial radius has also been calculated in lattice QCD~\cite{Green:2017keo,Alexandrou:2017hac,Capitani:2017qpc,Rajan:2017lxk}. However, these calculations suffer from different systematic errors and even adopt different methodologies to extract their uncertainties. A best estimate and a prescription for combining the errors from different studies, such as those performed by Ref.~\cite{Aoki:2016frl} for several other hadronic quantities, is still lacking. Therefore, for the following analysis, we take $\langle r_A^2 \rangle=[0.19,0.45]~\mathrm{fm}^2$, which covers the entire span of values along with the quoted uncertainties in Refs.~\cite{Green:2017keo,Alexandrou:2017hac,Capitani:2017qpc,Rajan:2017lxk}, as the lattice QCD result. In Fig.~\ref{fig:axial_ff}, we show the $\bar\nu/\nu$-$d$ cross sections with only 1B currents. For the range of kinematics shown here, the dipole parametrization with $M_A=1$~GeV gives cross sections that practically coincide with the model-independent expansion with $\langle r_A^2 \rangle=0.46~\mathrm{fm}^2$. Variations in the axial radius within the range of lattice QCD evaluations lead to 3-4\% uncertainty in the cross sections at $\epsilon\approx100$~MeV. The uncertainty estimates of Ref.~\cite{Hill:2017wgb} lead to 4-5\% variation in the cross sections at $\epsilon\approx100$~MeV, which are larger compared to the nuclear structure corrections discussed above and also compared to the size of the 2B current contributions. 
At $\epsilon\lesssim20$~MeV on the other hand, the NNLO$_\mathrm{sim}$ bands, which are larger than those from variation of the nucleon axial radius, provide a better  estimate of the total uncertainty of the calculation.

\subsection{The $\nopieft$ counterterm $L_{1,A}$}

 The low-energy regime lies well within the domain of validity of $\nopieft$,
which uses nucleons as the only dynamical degree of freedom. In contrast to $\chi$EFT, it can be applied in processes where the characteristic momentum $Q$ follows the scale hierarchy $Q~\approx~p,\,\gamma,\,1/a_s~\ll~m_\pi$, 
where $\gamma=45.701$~MeV is the deuteron binding momentum, and $a_s\approx-20$~fm is the NN s-wave scattering length in the spin-singlet channel. The EFT expansion is, therefore, in $Q/m_\pi$. At leading order in this expansion,  
$\nopieft$ provides $\bar\nu/\nu$-$d$ cross sections at $\epsilon\lesssim20$~MeV with a precision of 5-20\% in terms of $G_F$, $V_{ud}$, $g_A$ and NN scattering observables. The $\nopieft$ 2B currents which enter and next-to-leading order, however, contain three counterterms, $L_1$, $L_2$, and $L_{1,A}$, which need to be fixed by fitting to electroweak data. While $L_1$ and $L_2$ can be determined to high precision by fitting, for example, to experimental values of $np\rightarrow d\gamma$ rate and deuteron magnetic moment respectively, $L_{1,A}$ requires data from the weak sector. 
 Theoretical uncertainties of $\nopieft$ calculations of low-energy $\bar\nu/\nu$-$d$ scattering, like several other important weak processes such as proton-proton fusion, are  typically larger than the truncation error of their $\nopieft$  expansions due to the fact that the LEC $L_{1,A}$ has not been well determined.

Reference~\cite{Butler:2000zp} performed a next-to-next-to-leading order calculation of the $\bar\nu/\nu$-$d$ cross sections in terms of $a(\epsilon)$ and $b(\epsilon)$, where $\sigma(\epsilon) = a(\epsilon)+L_{1,A}\,b(\epsilon)$, with the renormalization scale $\mu$ set equal to the pion mass. Even though $a(\epsilon)$ and $b(\epsilon)$ were each calculated to better than 3\% precision for $\epsilon$ up to 20~MeV, $\sigma(\epsilon)$ could not be well constrained because $L_{1,A}$ was unknown. It was shown in Ref.~\cite{Kong:2000px} that the $\mu$ dependence of $L_{1,A}$ can be factorized out by writing 
\begin{equation}
\label{eq:l1a_rg}
 L_{1,A}=l_{1,A} \, 2 \pi \, g_A \, \frac{\sqrt{\rho_s \, \rho_t}}{(\mu-\gamma)\left(\mu-\frac{1}{a_s}\right)}\,,
\end{equation}
where $\rho_s=2.73$~fm is the NN effective range in the spin-singlet channel, whereas the spin-triplet (deuteron) channel effective range $\rho_t$ is 1.765~fm in the effective-range-expansion parametrization~\cite{Bethe:1949yr}, but is 2.979~fm in the zed parameterization~\cite{Phillips:1999hh}. The dimensionless coupling constant $l_{1,A}$ is independent of the renormalization scale.

By fitting the calculations of $\bar\nu/\nu$-$d$ scattering cross sections of Ref.~\cite{Butler:2000zp} to reactor antineutrino data, $L_{1,A}=3.6\pm5.5~\mathrm{fm}^3$ was obtained~\cite{Butler:2002cw}, whereas fitting with solar neutrino data at SNO gave $L_{1,A}=4.0\pm6.3~\mathrm{fm}^3$ ~\cite{Chen:2002pv}. The large uncertainties in both of these fits were due to statistical errors in the experiments. Apart from fitting to experimental data, LECs in EFTs can alternatively be determined by calculating them in the corresponding high-energy theory~\cite{Epelbaum:2008ga}. $L_{1,A}$ was recently computed directly in lattice QCD and the value $3.9(0.1)(1.0)(0.3)(0.9)~\mathrm{fm}^3$ was obtained~\cite{Shanahan:2017bgi}. 

In this work, we fit the calculations of Ref.~\cite{Butler:2000zp} to our $\chi$EFT results for $\sigma(\epsilon)$, which we treat as input data. To this end, we first 
update the $\nopieft$ results of Ref.~\cite{Butler:2000zp} for $a(\epsilon)\propto g_A^2$ and $b(\epsilon)\propto g_A$ to account for the updated value of the axial coupling constant from 1.26 used in Ref.~\cite{Butler:2000zp} to 1.2723~\cite{Tanabashi:2018oca} used in this work. It is important to note that the $\nopieft$ counterterm $L_{1,A}$ subsumes the effects of the pion-exchange axial currents and of the $\chi$EFT LEC $\hat d_R$. Therefore, the NNLO$_\mathrm{sim}$ constraints on the value of $L_{1,A}$, in essence, emerge from the fitting of $\hat c_{1,3,4}$ and $c_D$ along with all other $\chi$EFT LECs to selected $\pi$N and NN scattering data, energies and charge radii of $^{2,3}$H and $^{3}$He, the quadrupole moment of $^{2}$H, as well as $\beta$-decay width of $^{3}$H. This determination of $L_{1,A}$ is more systematic compared to the approach of Ref.~\cite{Butler:2000zp} that  fitted $L_{1,A}$ to phenomenological calculations in which the short-distance part of the axial 2B current was fixed by using $^{3}$H $\beta$ decay as input and that of Ref.~\cite{Mosconi:2002br} where it was fitted to calculations that modeled 2B currents as exchanges of pions and heavy bosons.

\begin{figure}[htbp]
\begin{center}
\includegraphics[width=\columnwidth,clip=true]{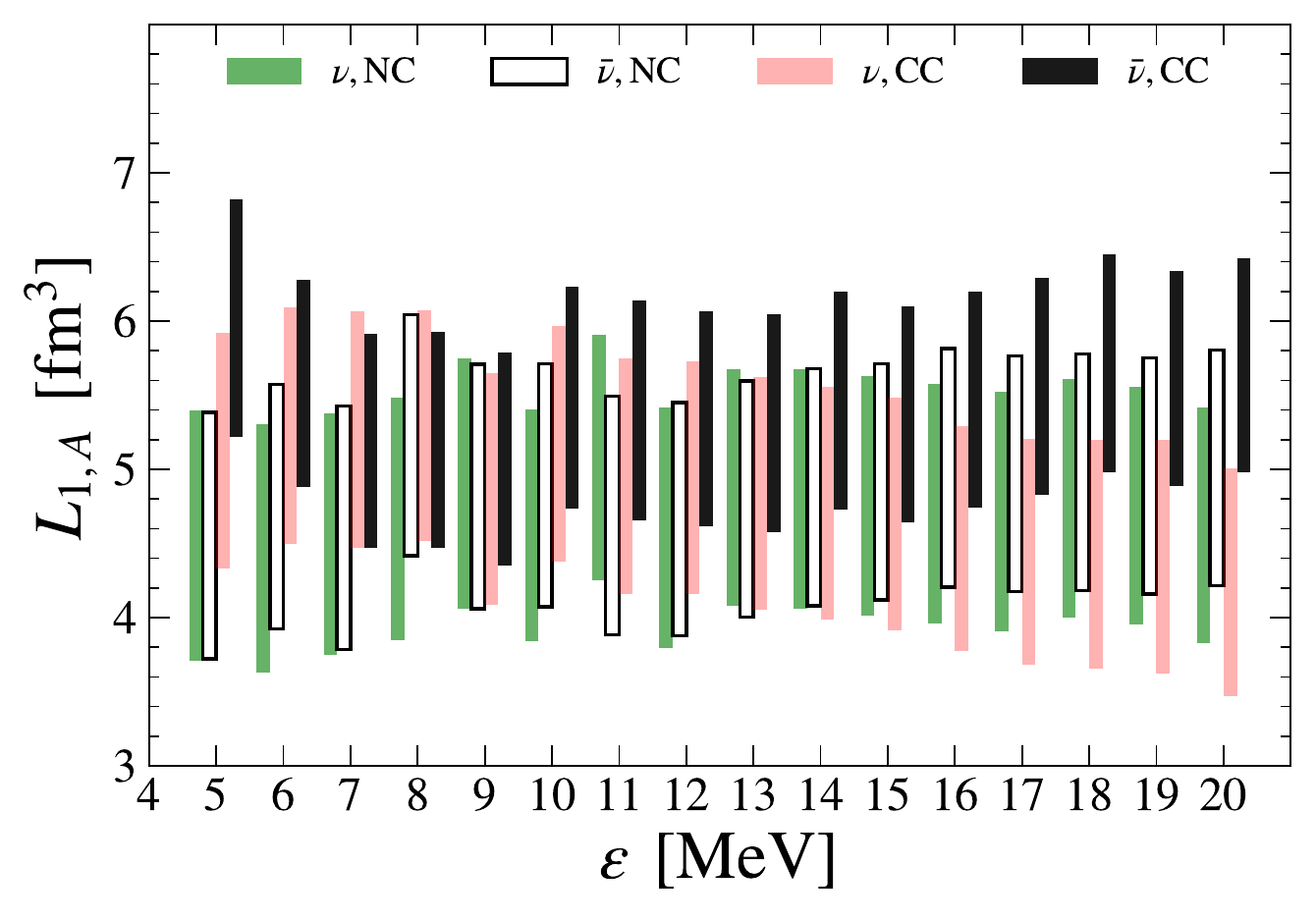} 
\end{center}
\caption{\protect (Color online) The $L_{1,A}$ values determined from $\nu$ NC (green/dark grey), $\bar\nu$ NC (white), $\nu$ CC (red/light grey), and $\bar\nu$ NC (black) processes. The vertical spreads of the bands are the NNLO$_\mathrm{sim}$ uncertainties. The $L_{1,A}$ values were calculated at 1~MeV intervals in the 5-20~MeV range of $\bar\nu$/$\nu$ energies, but have been slightly displaced along the horizontal axis for visibility.}
\label{fig:l1a}
\end{figure}

Figure~\ref{fig:l1a} shows the $L_{1,A}$ values
for the four reaction channels given by the NNLO$_\mathrm{sim}$ family of interactions.  Calculations are done on a grid of $1$ MeV in energy and are shown as bands that encompass the different values obtained with the 42 interactions for each of the four processes. One can clearly see that we get compatible constraints from all four processes. Averaging over the cross sections of all four channels, 16 energy values and the 42 interactions and using the spread of these values as a conservative uncertainty estimate, we obtain $L_{1,A}=4.9^{+1.9}_{-1.5}~\mathrm{fm}^3$. Although, in principle, one has to also add the EFT truncation uncertainties on $\sigma$, $a$ and $b$ in quadrature, their impact is negligible since they are much smaller ($\approx3\%$ each). 

Our value for $L_{1,A}$ is consistent with all of the above-mentioned determinations. Our constraint is narrower than those from $\bar\nu/\nu$-$d$ scattering experiments and is comparable with the lattice QCD result. The value $L_{1,A}=4.9^{+1.9}_{-1.5}~\mathrm{fm}^3$ corresponds to renormalization scale $\mu=m_\pi$. Using Eq.~\eqref{eq:l1a_rg}, we obtain $l_{1,A}=0.097^{+0.037}_{-0.029}$ using the effective-range-expansion parametrization of the NN scattering matrix in the deuteron channel and  $l_{1,A}=0.074^{+0.029}_{-0.023}$ in zed parametrization. The latter agrees with the value 0.051 obtained recently by Ref.~\cite{De-Leon:2016wyu} using a $\nopieft$ fit to $^3$H $\beta$-decay half-life, but the former does not agree with their corresponding value of 0.312. 

\section{Summary and Outlook}
\label{sec:conclusion}
We have developed an independent multipole decomposition framework to compute all of the four reaction channels of $\bar\nu/\nu$-$d$ inelastic scattering in $\chi$EFT. Our results agree with prior phenomenological and $\chi$EFT calculations.
We then perform an uncertainty quantification analysis of the four processes.
 Based on the observed convergence pattern of the $\chi$EFT expansion of the electroweak current operator and on the width of the NNLO$_\mathrm{sim}$ band which quantifies the short-distance model ambiguity of $\chi$EFT interactions, we estimate a nuclear structure uncertainty of about 3\% on the cross sections in the 100~MeV $\bar\nu/\nu$ energy region.

The large uncertainty in the recent lattice QCD calculations and phenomenological extractions of the axial radius renders it the dominant source of uncertainty  compared to nuclear structure uncertainties. This makes a precise determination of the axial nucleon form factor crucial for a high precision calculation of the deuteron cross section above 100 MeV in energy.
We expect the situation to be reversed in the neutrino cross section of heavier nuclei, where  nuclear structure uncertainty are typically larger 
due to the inherent complexity of the nuclear many-body problem and due to the presence of 3N forces. 

By matching our low-energy $\chi$EFT results to those of pionless effective field theory ($\nopieft$)~\cite{Butler:2000zp},  we provide a new constraint of the counterterm $L_{1,A}=4.9^{+1.9}_{-1.5}~\mathrm{fm}^3$ at $\mu=m_\pi$.
Our result is  consistent with a recent lattice QCD evaluation and narrower than prior experimental determinations from reactor antineutrino and solar neutrino data.   The uncertainty on $L_{1,A}$ is a major source of theory error on $\nopieft$ calculations of, e.g.,  the $S$ factor for the proton-proton fusion reaction which is  important in astrophysics~\cite{Chen:2012hm}. Our determination can therefore provide useful input for $\nopieft$ studies until a high precision experimental measurement~\cite{Andreev:2010wd} becomes available.

\section*{Acknowledgments}
We are thankful to Andreas Ekstr\"om for providing the NNLO$_\mathrm{sim}$ interactions, and to Wick Haxton, Daniel Phillips and Nir Barnea for fruitful discussions. This work was supported by the Cluster of Excellence ``Precision Physics, Fundamental  Interactions, and Structure of Matter (PRISMA$^+$),'' funded by  the German Research Foundation (DFG) within the German Excellence Strategy (Project ID 39083149), and by the DFG-funded Collaborative Research Center SFB 1044. We gratefully acknowledge the computing time granted on the supercomputer Mogon at Johannes Gutenberg-Universit\"at Mainz.


\begin{thebibliography}{99}

\bibitem{Benhar:2015wva} 
  O.~Benhar, P.~Huber, C.~Mariani and D.~Meloni,
  Phys.\ Rept.\  {\bf 700}, 1 (2017).
  
\bibitem{Nakamura:2000vp} 
  S.~Nakamura, T.~Sato, V.~P.~Gudkov and K.~Kubodera,
  Phys.\ Rev.\ C {\bf 63}, 034617 (2001); {\bf 73}, 049904 (2006).
  
\bibitem{Nakamura:2002jg} 
  S.~Nakamura, T.~Sato, S.~Ando, T.~S.~Park, F.~Myhrer, V.~P.~Gudkov and K.~Kubodera,
  Nucl.\ Phys.\ A {\bf 707}, 561 (2002).
  
\bibitem{Kubodera:1993rk} 
  K.~Kubodera and S.~Nozawa,
  Int.\ J.\ Mod.\ Phys.\ E {\bf 03}, 101 (1994).
  
\bibitem{Ahmad:2002jz} 
  Q.~R.~Ahmad {\it et al.} [SNO Collaboration],
  Phys.\ Rev.\ Lett.\  {\bf 89}, 011301 (2002).
  
\bibitem{Ahmed:2003kj} 
  S.~N.~Ahmed {\it et al.} [SNO Collaboration],
  Phys.\ Rev.\ Lett.\  {\bf 92}, 181301 (2004).

\bibitem{shen} G.~Shen, L.~E.~Marcucci, J.~Carlson, S.~Gandolfi and R.~Schiavilla, Phys.\ Rev.\ C {\bf 86}, 035503 (2012).
  
\bibitem{Gazit:2004sp} 
  D.~Gazit and N.~Barnea,
  Phys.\ Rev.\ C {\bf 70}, 048801 (2004).

\bibitem{Gazit:2007jt} 
  D.~Gazit and N.~Barnea,
  Phys.\ Rev.\ Lett.\  {\bf 98}, 192501 (2007).
  
\bibitem{OConnor:2007kup} 
  E.~O'Connor, D.~Gazit, C.~J.~Horowitz, A.~Schwenk and N.~Barnea,
  Phys.\ Rev.\ C {\bf 75}, 055803 (2007).
  
\bibitem{Golak:2018qya} 
  J.~Golak, R.~Skibi\'nski, K.~Topolnicki, H.~Wita\l{}a, A.~Grassi, H.~Kamada and L.~E.~Marcucci,
  Phys.\ Rev.\ C {\bf 98}, 015501 (2018).

\bibitem{Golak:2019fet} 
  J.~Golak, R.~Skibi\'nski, K.~Topolnicki, H.~Witala, A.~Grassi, H.~Kamada and L.~E.~Marcucci,
  Phys.\ Rev.\ C {\bf 100}, 064003 (2019).
  
\bibitem{Lovato:2015qka} 
  A.~Lovato, S.~Gandolfi, J.~Carlson, S.~C.~Pieper and R.~Schiavilla,
  Phys.\ Rev.\ C {\bf 91}, 062501 (2015).
  
\bibitem{Lovato:2014eva} 
  A.~Lovato, S.~Gandolfi, J.~Carlson, S.~C.~Pieper and R.~Schiavilla,
  Phys.\ Rev.\ Lett.\  {\bf 112}, 182502 (2014).
  
\bibitem{Lovato:2017cux} 
  A.~Lovato, S.~Gandolfi, J.~Carlson, E.~Lusk, S.~C.~Pieper and R.~Schiavilla,
  Phys.\ Rev.\ C {\bf 97}, 022502 (2018).
  
\bibitem{Rocco:2018mwt} 
  N.~Rocco, C.~Barbieri, O.~Benhar, A.~De Pace and A.~Lovato,
  Phys.\ Rev.\ C {\bf 99}, 025502 (2019).
  
\bibitem{Butler:2000zp} 
  M.~Butler, J.~W.~Chen and X.~Kong,
  Phys.\ Rev.\ C {\bf 63}, 035501 (2001).
  
\bibitem{Baroni:2017gtk} 
  A.~Baroni and R.~Schiavilla,
  Phys.\ Rev.\ C {\bf 96}, 014002 (2017).
  
\bibitem{Carlsson:2015vda} 
  B.~D.~Carlsson {\it et al.},
  Phys.\ Rev.\ X {\bf 6}, 011019 (2016).  
  
\bibitem{Andreas} A. Ekstr\"{o}m, private communication (2019).

\bibitem{walecka} J.~D.~Walecka, {\em Theoretical Nuclear and Subnuclear Physics}, 2nd ed. (World Scientific, Singapore and Imperial College Press, London, 2004).  

\bibitem{Weinberg:1990rz} 
  S.~Weinberg,
  Phys.\ Lett.\ B {\bf 251}, 288 (1990).
  
\bibitem{Weinberg:1991um} 
  S.~Weinberg,
  Nucl.\ Phys.\ B {\bf 363}, 3 (1991).
  
\bibitem{Epelbaum:1999dj} 
  E.~Epelbaum, W.~Gloeckle and U.-G.~Meissner,
  Nucl.\ Phys.\ A {\bf 671}, 295 (2000).
  
\bibitem{Entem:2003ft} 
  D.~R.~Entem and R.~Machleidt,
  Phys.\ Rev.\ C {\bf 68}, 041001 (2003).
  
\bibitem{Epelbaum:2008ga} 
  E.~Epelbaum, H.~W.~Hammer and U.-G.~Meissner,
  Rev.\ Mod.\ Phys.\  {\bf 81}, 1773 (2009).
  
\bibitem{Machleidt:2011zz} 
  R.~Machleidt and D.~R.~Entem,
  Phys.\ Rept.\  {\bf 503}, 1 (2011).

\bibitem{Krebs:2019uvm} 
  H.~Krebs,
  arXiv:1908.01538 [nucl-th].
  
\bibitem{Feenberg:1950vj} 
  E.~Feenberg and G.~Trigg,
  Rev. \ Mod. \ Phys.\ {\bf 22}, 399 (1950).

\bibitem{VMK} D.A.~Varshalovich, A.N.~Moskalev, V.K. Khersonskii, ``Quantum Theory of Angular Momentum'' (World Scientific, 1988).


  
\bibitem{Schiavilla:2004wn} 
  R.~Schiavilla, J.~Carlson and M.~W.~Paris,
  Phys.\ Rev.\ C {\bf 70}, 044007 (2004).

  
\bibitem{Carlson:2001ma} 
  J.~Carlson, R.~Schiavilla, V.~R.~Brown and B.~F.~Gibson,
  Phys.\ Rev.\ C {\bf 65}, 035502 (2002).
  
\bibitem{Park:1995pn} 
  T.~S.~Park, D.~P.~Min and M.~Rho,
  Nucl.\ Phys.\ A {\bf 596}, 515 (1996).

\bibitem{park_axial} T.~S.~Park, L.~E.~Marcucci, R.~Schiavilla, M.~Viviani {\em et al.}, Phys.\ Rev.\ C {\bf 67}, 055206 (2003).

\bibitem{phillips} D.~R.~Phillips, Phys.\ Lett.\ B {\bf 567}, 12 (2003).

\bibitem{Kolling:2009iq} 
  S.~K\"olling, E.~Epelbaum, H.~Krebs and U.-G.~Meissner,
  Phys.\ Rev.\ C {\bf 80}, 045502 (2009).
  
\bibitem{Kolling:2011mt} 
  S.~K\"olling, E.~Epelbaum, H.~Krebs and U.-G.~Meissner,
  Phys.\ Rev.\ C {\bf 84}, 054008 (2011).

\bibitem{Krebs:2016rqz} 
  H.~Krebs, E.~Epelbaum and U.-G.~Meissner,
  Ann. Phys.\ (N.Y.)\ {\bf 378}, 317 (2017).
  
\bibitem{Krebs:2019aka} 
  H.~Krebs, E.~Epelbaum and U.-G.~Meissner,
  Few Body Syst.\ {\bf 60}, 31 (2019).
  
\bibitem{Pastore:2008ui} 
  S.~Pastore, R.~Schiavilla and J.~L.~Goity,
  Phys.\ Rev.\ C {\bf 78}, 064002 (2008).
  
\bibitem{Pastore:2009is} 
  S.~Pastore, L.~Girlanda, R.~Schiavilla, M.~Viviani and R.~B.~Wiringa,
  Phys.\ Rev.\ C {\bf 80}, 034004 (2009).
  
\bibitem{Pastore:2011ip} 
  S.~Pastore, L.~Girlanda, R.~Schiavilla and M.~Viviani,
  Phys.\ Rev.\ C {\bf 84}, 024001 (2011).
  
\bibitem{Baroni:2015uza} 
  A.~Baroni, L.~Girlanda, S.~Pastore, R.~Schiavilla and M.~Viviani,
  Phys.\ Rev.\ C {\bf 93}, 015501 (2016); {\bf 93}, 049902 (2016); {\bf 95}, 059901 (2017).
  
\bibitem{Phillips:2016mov} 
  D.~R.~Phillips,
  Annu.\ Rev.\ Nucl.\ Part.\ Sci.\  {\bf 66}, 421 (2016).
  
\bibitem{Kubis:2000zd} 
  B.~Kubis and U.-G.~Meissner,
  Nucl.\ Phys.\ A {\bf 679}, 698 (2001).

\bibitem{Bhattacharya:2011ah} 
  B.~Bhattacharya, R.~J.~Hill and G.~Paz,
  Phys.\ Rev.\ D {\bf 84}, 073006 (2011).
    
\bibitem{Bernard:1994wn} 
  V.~Bernard, N.~Kaiser and U.-G.~Meissner,
  Phys.\ Rev.\ D {\bf 50}, 6899 (1994).
  
\bibitem{Gazit:2008ma} 
  D.~Gazit, S.~Quaglioni and P.~Navratil,
  Phys.\ Rev.\ Lett.\  {\bf 103}, 102502 (2009); {\bf 122}, 029901 (2019). 

  
\bibitem{Andreev:2010wd} 
  V.~A.~Andreev {\it et al.} [MuSun Collaboration],
  arXiv:1004.1754 [nucl-ex].

\bibitem{Acharya:2018qzk} 
  B.~Acharya, A.~Ekstr\"om and L.~Platter,
  Phys.\ Rev.\ C {\bf 98}, 065506 (2018).
  
\bibitem{Wiringa:1994wb} 
  R.~B.~Wiringa, V.~G.~J.~Stoks and R.~Schiavilla,
  Phys.\ Rev.\ C {\bf 51}, 38 (1995).
  
\bibitem{Meyer:2016oeg} 
  A.~S.~Meyer, M.~Betancourt, R.~Gran and R.~J.~Hill,
  Phys.\ Rev.\ D {\bf 93}, 113015 (2016).
  
\bibitem{Singh:1971md} 
  S.~K.~Singh,
  Nucl.\ Phys.\ B {\bf 36}, 419 (1972).
  
\bibitem{Hill:2017wgb} 
  R.~J.~Hill, P.~Kammel, W.~J.~Marciano and A.~Sirlin,
  Rept.\ Prog.\ Phys.\ {\bf 81}, 096301 (2018).

\bibitem{Green:2017keo} 
  J.~Green {\it et al.},
  Phys.\ Rev.\ D {\bf 95}, 114502 (2017).
  
\bibitem{Alexandrou:2017hac} 
  C.~Alexandrou, M.~Constantinou, K.~Hadjiyiannakou, K.~Jansen, C.~Kallidonis, G.~Koutsou and A.~Vaquero Aviles-Casco,
  Phys.\ Rev.\ D {\bf 96}, 054507 (2017).
  
\bibitem{Capitani:2017qpc} 
  S.~Capitani {\it et al.},
  Int.\ J.\ Mod.\ Phys.\ A {\bf 34}, 1950009 (2019).
  
\bibitem{Rajan:2017lxk} 
  R.~Gupta, Y.~C.~Jang, H.~W.~Lin, B.~Yoon and T.~Bhattacharya,
  Phys.\ Rev.\ D {\bf 96}, 114503 (2017).
  
\bibitem{Aoki:2016frl} 
  S.~Aoki {\it et al.},
  Eur.\ Phys.\ J.\ C {\bf 77}, 112 (2017).
  
\bibitem{Kong:2000px} 
  X.~Kong and F.~Ravndal,
  Phys.\ Rev.\ C {\bf 64}, 044002 (2001).
  
\bibitem{Bethe:1949yr} 
  H.~A.~Bethe,
  Phys.\ Rev.\  {\bf 76}, 38 (1949).
  
\bibitem{Phillips:1999hh} 
  D.~R.~Phillips, G.~Rupak and M.~J.~Savage,
  Phys.\ Lett.\ B {\bf 473}, 209 (2000).

\bibitem{Butler:2002cw} 
  M.~Butler, J.~W.~Chen and P.~Vogel,
  Phys.\ Lett.\ B {\bf 549}, 26 (2002).

\bibitem{Chen:2002pv} 
  J.~W.~Chen, K.~M.~Heeger and R.~G.~Hamish~Robertson,
  Phys.\ Rev.\ C {\bf 67}, 025801 (2003).
  
\bibitem{Shanahan:2017bgi} 
  P.~E.~Shanahan {\it et al.},
  Phys.\ Rev.\ Lett.\  {\bf 119}, 062003 (2017).
  
\bibitem{Tanabashi:2018oca} 
  M.~Tanabashi {\it et al.} [Particle Data Group],
  Phys.\ Rev.\ D {\bf 98}, 030001 (2018).
  
\bibitem{Mosconi:2002br} 
  B.~Mosconi, P.~Ricci and E.~Truhl\'ik,
  Nucl.\ Phys.\ A {\bf 772}, 81 (2006).
  
\bibitem{De-Leon:2016wyu} 
  H.~De-Leon, L.~Platter and D.~Gazit,
  Phys.\ Rev.\ C {\bf 100}, 055502 (2019).
  
\bibitem{Chen:2012hm} 
  J.~W.~Chen, C.-P.~Liu and S.~H.~Yu,
  Phys.\ Lett.\ B {\bf 720}, 385 (2013).
  
\end{thebibliography}
\end{document}